\documentstyle[aps,eqsecnum,psfig,preprint]{revtex}
\global\firstfigfalse
\newcommand\email[1]{\footnote{#1}}

\def\bea{\begin{eqnarray}}
\def\eea{\end{eqnarray}}
\def\bce{\begin{centering}}
\def\ece{\end{centering}}
\def\bit{\begin{itemize}}
\def\eit{\end{itemize}}
\def\none{\widetilde N_1}
\newcommand\n[1]{\widetilde N_{#1}}
\def\grav{\widetilde G}
\def\bino{\widetilde B}
\def\lsp{\widetilde\psi}
\def\cone{\widetilde C_1}
\def\ntwo{\widetilde N_2}
\def\nthree{\widetilde N_3}
\def\stau{\widetilde\tau}
\def\snut{\widetilde\nu_\tau}
\def\xvis{x_{vis}}
\def\att{\widetilde{A}_{\tau}}
\def\DM{\Delta M}
\def\tightenlines{\def\baselinestretch{1.2}\small\normalsize}
\def\slashchar#1{\setbox0=\hbox{$#1$}           
  \dimen0=\wd0                                 
  \setbox1=\hbox{/} \dimen1=\wd1               
  \ifdim\dimen0>\dimen1                        
  \rlap{\hbox to \dimen0{\hfil/\hfil}}      
  #1                                        
  \else                                        
  \rlap{\hbox to \dimen1{\hfil$#1$\hfil}}   
  /                                         
  \fi}                    
\def\met{\slashchar{E}_T}                     
\tightenlines

\begin{document}
\draft
\preprint{UCD-2000-14}
\title{How to Interpret a Tau Excess at LEP2 within the MSSM}
\author{S. Mrenna\email{mrenna@physics.ucdavis.edu}\\
\it Physics Department, University of California at Davis, Davis, CA  95616, USA}

\date{\today}
\maketitle

\begin{abstract}
Neutralino and tau slepton pair production can naturally
produce an excess of tau lepton pairs at the current 
LEP collider energies.
We describe the constraints this has on the 
values of the mass parameters in the softly
broken Supersymmetric Lagrangian, and
consider the consequences for
superpartner production at LEP and at the Fermilab Tevatron collider.
The pair production of the LSP and a heavier neutralino,
followed by a 2-body decay to a tau slepton and tau lepton,
is consistent with the present LEP data, predicts
a chargino mass below 125 GeV,
and provides an interesting Cold Dark Matter component,
with $\Omega h^2\sim .1-.2$. 
\end{abstract}

\newpage

\section {Introduction}

Recent measurements at LEP suggest
an excess of events containing pairs of tau leptons, though the
observation is only of marginal statistical significance \cite{lepc}:
around $3\sigma$ ignoring systematic errors.
It is interesting to study such a possibility, even if only to
prepare oneself for the future when the first definitive hint of
physics beyond the Standard Model emerges.
An excess of tau lepton pairs could arise 
from the direct production of a pair of tau 
sleptons,
which subsequently decay to tau leptons and gravitino or neutralino
LSP's.   The tau sleptons may also be produced indirectly in
the decays of charginos or neutralinos.
Using the available 
data, a kinematic and dynamical analysis is performed to exclude
some mechanisms and constrain others.  Based on this, one can more easily 
establish the existence of a small signal if it is there.  
We map out the softly broken SUSY Lagrangian,
show how the mass and wave function of the LSP
can be deduced from kinematic distributions and event rates,
and study the implications for the Fermilab Tevatron.

\section{The data and its SUSY interpretation}

An overall excess of tau leptons is observed in all four
LEP experiments in a sample of 220 pb$^{-1}$ of data 
per experiment
accumulated between $\sqrt{s}=2E_{beam}=192$ and $202$ GeV.  
There appears to be a threshold above $\sqrt{s}=183$ GeV, possibly
above $\sqrt{s}=189$ GeV, with
a steady excess distributed over energy.  
The excess 
sets the size of the production cross section times branching
ratio times detection efficiency at about $.03$ pb. 
The kinematics are indicative of a large mass splitting $\DM$ 
between a heavier superpartner and the LSP.  No other excesses
are reported by all four experiments. 
Our objective is
to take this observation seriously and consider its SUSY
implications.

The experimental analysis assumes
a specific SUSY signal 
with a range of $\DM$ values, and cuts on the observables are
optimized for this signal.  
There is no direct measurement of $\DM$.
Instead, quantities such as the visible energy 
$E_{vis}$ are measured which are correlated with $\DM$.
In the remainder, we will focus on the fraction of visible
energy $x_{vis}\equiv E_{vis}/\sqrt{s}$ 
as the main characteristic of the signal.
The $\Delta M$
range depends on the hypothesized SUSY signal, and thus is
different for slepton pair production or neutralino/chargino
production.  For slepton pair production, the L3 experiment
analyzes the low, medium, and
high $\Delta M$ ranges defined 
as $5-20$, $20-40$, and $40^+$ GeV \cite{Acciarri:1999kk}.
For tau slepton production, the maximum value in each range is
degraded somewhat, since tau decays have an intrinsic
missing energy.
For neutralino or chargino pair production, the upper end of each
range is roughly double those for slepton pair production \cite{Acciarri:2000km}.
This can be understood roughly as the difference between
the kinematics of 2- and 3-body decays.

The optimized cuts used by L3 for the large $\DM$ signal of
the tau slepton analysis can be summarized as
two reconstructed clusters of particles
satisfying the following requirements \cite{Acciarri:1999kk}:
\begin{eqnarray}
.1\le {x_{vis}} \le .4 ~~~~~
.05\le {p_T \over \sqrt{s}} \le .3 \nonumber \\
E^{\ell}/E_{beam} < {1\over 3} ~~~~~
\sin\slashchar\theta > 0.55 ~~~~~
\Delta\phi_{\tau_1\tau_2}<2.77.
\label{eq:cuts}
\end{eqnarray}
The experimental quantities used above are the
summed $p_T$ of the two clusters, the electron or muon energy $E^{\ell}$,
the polar angle $\slashchar\theta$ of the missing energy vector
$\slashchar{p}^\mu=p^\mu_{e^+}+p^\mu_{e-}-p^\mu_{1}-p^\mu_2$ 
where $1$ and $2$ denote the two final
state clusters, and the azimuthal angle between the two clusters
$\Delta\phi_{\tau_1\tau_2}$.
When a quantitative measure of the efficiency is needed,
the signature $e$ or $\mu$+low-multiplicity hadronic cluster is used,
since it represents a large branching fraction with little QCD background.
These cuts have an efficiency of approximately $0.3$ on an hypothesized
signal of tau slepton pair production.

\begin{figure}[ht] 
\centerline{\psfig{file=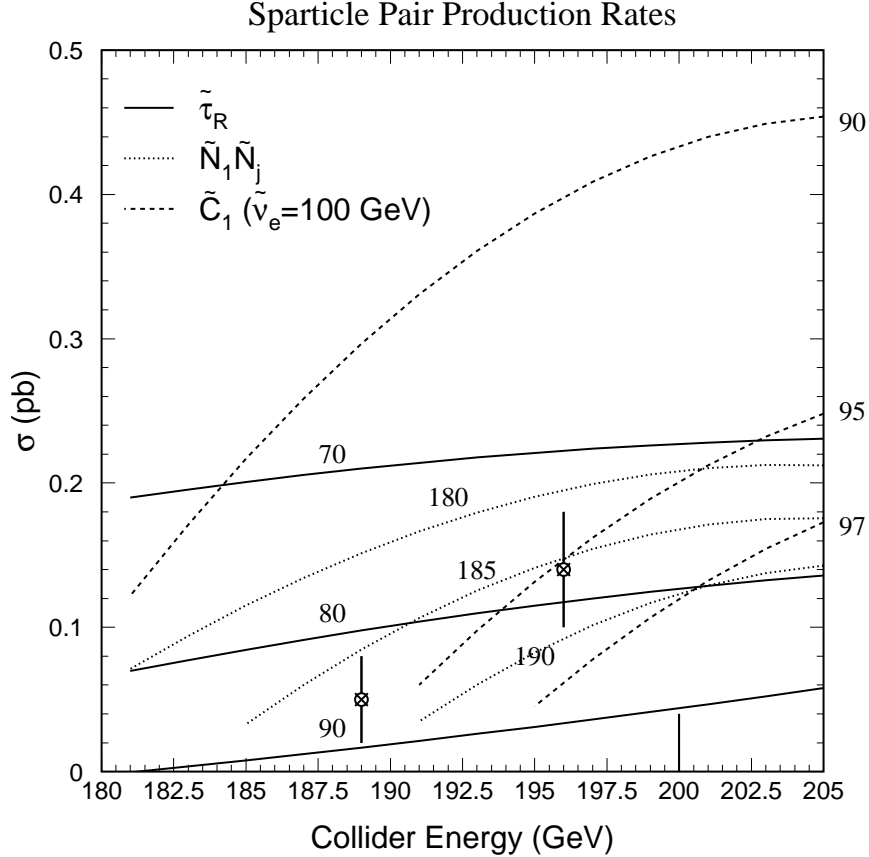,width=12cm}}
\caption{Sparticle pair production rates from $e^+e^-$ annihilation
as a function of center-of-mass energy.  The data points
are preliminary results of the LEP SUSY Working Group.}
\label{fig:xsections}
\end{figure}

Our starting hypothesis for a phenomenological analysis
is that the tau excess results
from the production of tau slepton, neutralino, or
chargino pairs or some combination of these.
For reference, various sparticle production cross sections are 
shown as a function of collider energy in
Figure~\ref{fig:xsections}.  The sum of the $\none\ntwo$
and $\none\nthree$ processes is shown 
for the cases of $M_{\none}+M_{\nthree}=180, 185$ and $190$ GeV (short-dashed lines),
the $\stau_R\stau^*_R$ process for $M_{\stau_R}=80, 85$, and 90 GeV (solid lines),
and the $\cone^+\cone^-$ process for $M_{\cone}=90, 95,$ and 97 GeV (long-dashed lines).
The cross sections were 
calculated using \protect{PYTHIA \cite{spythia}}, with
$M_1={1\over 2}\mu>0, \tan\beta=5,$ 
$M_2=M_{\tilde{e}_L}=M_{\tilde{e}_R}=$300 GeV for neutralino
pair production, and
 $\mu=-1$ TeV, $\tan\beta=5$, $M_1=40$ GeV, and $M_{\tilde{\nu}_e}=100$ 
GeV for chargino pair production.

The data points, denoted by
the circles with error bars \cite{favara}, are from a preliminary
analysis by the LEP SUSY Working Group assuming tau slepton pair
production.  
The data appears to be compatible with direct tau slepton
production and $M_{\stau}\simeq 80$ GeV or with
$\none\n j$ production with $M_{\none}+M_{\nthree}=$ 185 GeV
and a heavier tau slepton.  Because of its quick rise with energy,
the chargino cross section does not agree well with the data.
In the next few sections, we fill in the details surrounding these
statements.
Since neutralino pair production is an indirect source of 
tau sleptons, we describe first the simpler case of direct
stau pair production.

\subsection{Tau slepton pair production}
The production and decay chain
$e^+e^-\to \stau_1(\to \tau^-\lsp) \stau_1^*(\to\tau^+\lsp)$,
where $\stau_1$ is the lightest tau slepton and $\lsp$ may be a
light gravitino or the lightest neutralino, can produce an excess
of tau leptons with missing energy.
For simplicity of the discussion, it is assumed that the $\stau_1$ is 
a pure interaction eigenstate, either $\stau_L$ or $\stau_R$.  
Tau slepton pairs are produced through $\gamma^*$ and $Z^*$ decays.
The $Z^*$ contribution depends on the tau eigenstate, but this effect is
small.
The $\stau_L$ cross section is slightly larger than the
$\stau_R$ one, and in the limit that $\sin^2\theta_W={1\over 4}$,
the following relation holds:
\begin{eqnarray}
{\sigma(\stau_L\stau_L)\over\sigma(\stau_R\stau_R)} =
\left(1+{1\over 9(1-M_Z^2/s)^2}\right),
\label{eq:ratio}
\end{eqnarray}
and there is little variation in the total rate from left-right mixing.
To get the desired cross section, the
tau slepton must be fairly light, but it also must have avoided
detection at the $Z$ pole and at other intermediate collider
energies.  
From Fig.~\ref{fig:xsections}, it can be seen that
a right-handed stau with a mass of about 80 GeV yields the
correct cross section, assuming that the efficiency is about $0.3$.
Mixing cannot make the cross section arbitrarily small, so a light
tau slepton is already excluded.

As mentioned previously, a large fraction of visible energy is the main
characteristic of the signal.
An upper bound on the visible energy in each event 
is roughly the sum of the tau energies.  
In the rest $(*)$ frame of the tau slepton, the tau energy is
$E_\tau^*=(M^2_{\stau}-M^2_{\lsp})/(2M_{\stau})$.
The boost to the lab frame introduces a factor $\gamma=E_{beam}/M_{\stau}$,
and the fraction of visible energy is:
\begin{eqnarray}
x_{vis}\equiv {E_{vis}\over\sqrt{s}}\simeq {2E_\tau\over\sqrt{s}}=
{1\over 2}\left(1-{M^2_{\lsp} \over M^2_{\stau}} \right).
\end{eqnarray}
This estimate ignores the direction of the boost and the energy
lost to neutrinos in $\tau$ decays, but is useful in 
distinguishing models.
The value of $1/2$ is obtained for a light gravitino LSP, which
yields a large fraction of
visible energy.  However, we would
have arrived at the same estimate for $W$ pair production, which
is the dominant background, so one should be careful about drawing
conclusions about the size of a signal before considering the
cuts necessary to reduce the backgrounds.
For the case of a neutralino LSP, a fraction $.3(.4)$ 
requires $M_{\none}/M_{\stau}=.64 (.45)$.  This illustrates the
necessary mass relations to explain the kinematics of the data.

For $\DM=M_{\stau}-M_{\lsp}>30$ GeV, 
the efficiency after the cuts listed in 
Eq.~\ref{eq:cuts} is 
roughly constant at $0.3$ (see Table 3 of Ref.~\cite{Acciarri:1999kk}.
The efficiencies listed there are in accord with our own simulations).
Therefore, a NLSP $\stau_1 = \stau_R$ with $M_{\stau}\sim 80$ GeV and
a large mass splitting between the tau slepton and the LSP can
explain the tau excess.
For a neutralino LSP, this predicts $36<M_{\none}<51$ GeV for
$.3<\xvis<.4$.
There is no rigorous experimental lower bound on $M_{\none}$ if
the $\none$ wave function contains no Higgsino components.
Since the kinematics suggest $M_{\none} < M_Z/2$, and a light
Wino- or Higgsino-like LSP requires a light chargino, a neutralino 
LSP should be Bino-like.

If $\stau_1=\stau_L$, in which case the stau is no longer the
NLSP, a light $\snut$ is
also required by the sum rule $M^2_{\stau_L}-M^2_{\snut}=-\cos{2\beta}M_W^2$.
Once $\tan\beta$ is larger than a few, it is a good approximation to take the squared
mass splitting to be $M_W^2$.  The lower bound on the sneutrino mass is
$M_Z/2$, leading to the bounds $M_{\stau}\ge 91$ GeV.
From Fig.~\ref{fig:xsections}, one can estimate that
$\stau_1=\stau_L$ can be marginally consistent with the data.
If one does not assume large $\tan\beta$, then $\tan\beta>2.31$ allows
for $M_{\snut}>M_Z/2$ and $M_{\stau_L}=80$ GeV.

Apropos to a $\stau_L$ solution,
if the gravitino is the LSP and  $M_{\grav}$ is heavier than about a meV, the decay
$\stau_L\to \snut W^*$ will occur before $\stau_L\to\tau\grav$, because of the
small gravitino coupling to matter.
This is not compatible with the data, because of the many $W^*$ decay modes.
Therefore, a gravitino LSP and a light $\stau_L$ can be rejected.
The possibility that $\widetilde\nu_\tau$ is the LSP and $\none$
is heavy can also be rejected, since the
decays $\stau_L\to W^*\snut$ would also occur.
On the other hand, 
the mass hierarchies $M_{\stau_L}> M_{\none}> M_{\snut}$ and
$M_{\stau_L}>M_{\snut}>M_{\none}$ are viable.  For the former case,
$M_{\none}$ must be close to $M_{\snut}$ to satisfy the kinematic
requirement $M_{\none}/M_{\stau}< .45-.64$, 
while the latter case allows for a substantially
lighter $\none$.  The existence of $\snut$ with $M_{\snut}>M_Z/2$ is
not very relevant, as long as $\stau_L\to\tau\none$ is the dominant
stau decay, since $\none\to\nu\snut$ and $\snut\to\nu\none$ are
both invisible decays, and direct $\none\none$ LSP and $\snut\snut^*$ LSP 
production are difficult to observe.

An admixture of $\stau_L$ and $\stau_R$ pair production
is possible, and this would lead to mixed kinematics in the data.
Kinematic correlations between the tau pair decay products can provide
more information on the quantum numbers of the tau slepton.
The tau lepton 
carries $L$ or $R$ polarization (in the rest frame of the stau) 
depending on whether the parent is $\stau_L$ or $\stau_R$.  
Pair production of $\stau_R$ then yields a correlation in the
kinematics of the decay products of the tau leptons.  As an
example, if the decays $\tau_R^-\to\pi^-\nu_\tau$ and $\tau_R^+
\to\pi^+\bar{\nu}_{\tau}$ occur, the $\pi^-$ will tend to be
hard while the $\pi^+$ is soft, {\it i.e.} the correlation is
$RR\to\pi^-_h\pi^+_s$ or $LL\to\pi^-_s\pi^+_h$ where $RR$ and $LL$
denote the stau wave function and a subscript $s$ or $h$ denotes
soft or hard kinematics.  These considerations are relevant even
if the tau sleptons are produced through on-shell decays of
neutralinos or charginos.  This polarization effect is included
in our particle-level simulations.

\begin{figure}[!h]
\centerline{\psfig{file=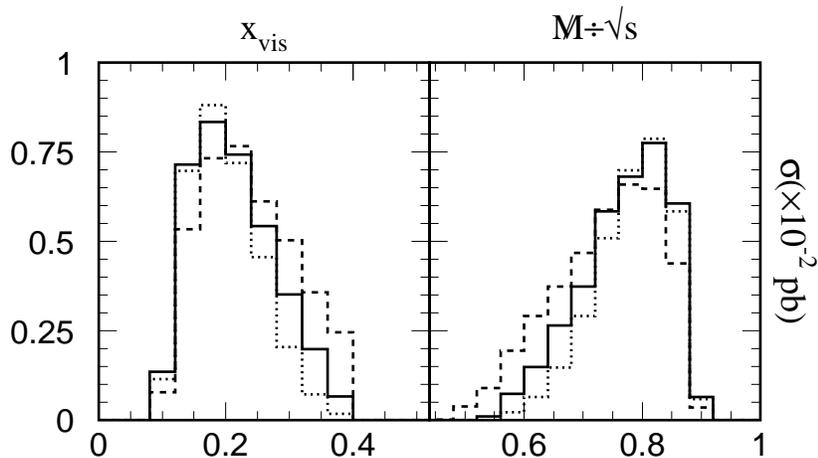,width=12cm}}
\caption{Kinematic distributions from tau slepton pair production
assuming different LSP's and from $\none\nthree$ production with
a $\none$ LSP.  The dominant $WW$ background is not included.}
\label{fig:stau_dist}
\end{figure}

To summarize, a tau excess can be explained by 
tau slepton pair production for a narrow range of tau slepton
masses.  
The cross section (using an efficiency of $.3$) is compatible with
a $\stau_R$ with mass around 80 GeV.  The stau may be left-handed,
but large $\tan\beta$ requires $M_{\stau_L}>90$ GeV.
The kinematics of the decays suggest
$M_{\lsp}/M_{\stau}< .45-.64$, so that
that $\lsp=$ $\grav$ or $\bino$ is possible.
The two LSP cases are kinematically distinguishable, or,
more precisely, the mass of the LSP can be inferred from kinematic
distributions.
This is 
illustrated in Fig.~\ref{fig:stau_dist}, 
which shows the $E_{vis}$ and $\slashchar{M}$ distributions
divided by $\sqrt{s}=200$ GeV
for $M_{\stau_R}=80$ GeV
and a neutralino LSP with $M_{\none}=36$ GeV (solid lines)
and a gravitino LSP (long-dashed  lines).
The distributions from $\none\nthree$ production are also shown
(short-dashed lines) for $M_{\none}+M_{\nthree}=185$ GeV and
$M_{\stau}=90$ GeV.  The dominant $WW$ background is not included.
A measurement of the cross section at several energies, as well
as analysis of kinematic distributions such as these, would allow
the various scenarios to be separated.

\subsection{Neutralino Pair Production}
The production and decay chain
$e^+e^-\to\none\n j$, $\n j\to\stau\tau$,
$\stau\to\tau\lsp$, where $\lsp$ is the LSP and $j=2$ and/or $3$, 
can also yield
a tau pair and missing energy.  The case of $\lsp=\none$ is considered
here with a brief comment on a gravitino LSP.
The relevance of
3-body decays of the neutralino is also addressed.
The $\none\n j$ process can be
kinematically accessible before chargino and possibly tau slepton
pair production.
The kinematic requirements set by the data are that 
$M_{\n j}-M_{\none}\sim 80$ GeV, while the production cross section
requires $183<M_{\none}+M_{\n j}<189$ GeV,
so that $M_{\n j}\sim 135, M_{\none}\sim 55$ GeV is 
marginally plausible.
Of course, it is necessary that $M_{\n j}-M_{\none}<M_Z$ to prevent
$\n j\to\none Z$ as the dominant decay.

The production of neutralino pairs proceeds
through the $Z$ in the $s$-channel or through $t$-channel exchanges
of selectrons. 
In the notation of Haber and Kane \cite{Haber:1985rc}, 
the $Z\none\n j$ coupling is
$O^{''L}_{1j}=-{1\over 2}N_{13}N^*_{j3}+{1\over 2}N_{14}N^*_{j4}$.
For $\none\n j$ production to be relevant, both neutralinos must
have some Higgsino component.
This requirement means that the $Z^*$ contribution to the decay can
be significant if 2-body decays are not allowed.
Because the excess appears in tau lepton final states,
we assume the selectron is not too light.

Since the $\none$ wave function should contain some Higgsino
component, it is necessary to
consider the LSP contribution to the invisible width of the $Z$.
We require that $\Gamma(Z\to\none\none)$ 
is less than 3.0 MeV \cite{Caso:1998tx} when $M_{\none}<M_Z/2$,
where the $\none\none$ contribution is:
\begin{eqnarray}
\Gamma(Z\to\none\none)={\alpha M_Z\beta \over 6\cos^2\theta_W\sin^2\theta_W}
\left[|O^{''L}_{11}|^2(1-r^2)-3r^2 Re((O^{''L}_{11})^2)\right] \nonumber \\
r={M_{\none}\over
M_Z} ~~~~~ \beta=\sqrt{1-4r^2},
\end{eqnarray}
with $O^{''L}_{11}=-{1\over 2}N_{13}N^*_{13}+{1\over 2}N_{14}N^*_{14}$.

Two-body decays of $\n j$ can satisfy the kinematic
requirements if there is 
sufficient mass splitting $M_{\n j}-M_{\stau}$ so that the tau produced
in the cascade $\n j\to\stau\tau$ is not too soft.  
As before, it is useful to have a simple estimate of the 
kinematics $\xvis$.
Again, the visible energy is approximately the sum of the tau energies.
Since the tau leptons
result from the cascade decays of heavy objects, $\n j\to\tau\stau,
\stau\to\tau\none$, the tau energies can be approximated by their
values in the rest frame of their individual parents.  
This assumes that the kinematic boost is small.
The fraction of visible energy can then be described approximately by:
\begin{eqnarray}
{x_{vis}} \simeq {1\over\sqrt{s}}
\left( {M^2_{\ntwo}-M^2_{\stau} \over 2 M_{\ntwo}} +
       {M^2_{\stau}-M^2_{\none} \over 2 M_{\stau}} \right).
\end{eqnarray}
The example noted above, $M_{\n j}\sim 135$ GeV,
$M_{\stau}\sim 95-100$ GeV, $M_{\none}\sim 55$ GeV, can satisfy the
nominal kinematic requirements, with $\xvis\simeq 0.3$.  
Because of the $p$-wave suppression
of the production cross section, $\stau\stau$ production would be
negligible in this situation.  

The example cross sections shown in Figure~\ref{fig:xsections} for neutralino
pair production are within range of the observed tau excess, while large
values of $\xvis$ can be obtained for sufficiently large mass splittings.
A scan over a range of $M_1, M_2, \mu$ and $\tan\beta$ revealed solutions
of the form:
\begin{eqnarray}
183 < M_{\none}+M_{\ntwo} \le 190 {\rm~GeV}, ~~~~~ x_{vis}\ge .3,
~~~~~ M_{\cone} \ge 100~{\rm GeV}, ~~~~~ \Gamma(Z\to\none\none)<3~{\rm MeV}, 
\nonumber \\
\sigma(\sqrt{s}=192~{\rm GeV})\times\epsilon \ge .03~{\rm pb}, ~~~~~ 
\sigma(\sqrt{s}=200~{\rm GeV})\times\epsilon \ge .03-.06~{\rm pb}, 
\end{eqnarray}
where $\sigma$ is the total sparticle production cross section
and $\epsilon$ is the detection efficiency using the cuts
described in Eq.~\ref{eq:cuts}.
For each case, $M_{\stau}$ was chosen to maximize $\xvis$.
The efficiencies $\epsilon$ from particle level simulations are
approximately $0.3$, similar to those obtained for tau slepton pair
production.
To decrease the size of the production cross section
$\sigma(\none\n j)$ to a plausible level, the selectron
mass parameters were both fixed at 300 GeV.  
The requirement of a heavy chargino not only prevents chargino
pair production, but also suppresses the decay $\n j\to\cone f\bar f'$.

The observed cross section
$\sigma\times\epsilon$ is allowed to be as large as $.06$ pb, since there 
is some flexibility in setting the overall rate.
A heavier selectron mass will significantly
reduce the $\none\n j$ cross section.  Tau slepton pair production
also contributes to the total event rate, and left-right mixing
can reduce the $\stau\stau$ contribution, or the tau slepton mass
can be larger than the value with maximizes the 
estimate $\xvis$, thereby reducing the cross section.  
If $\stau_L$ is light,  
instead of $\stau_R$, then the decay $\n j\to \snut\bar\nu_\tau\to\nu_\tau\bar\nu_\tau\none$ will reduce the visible cross section.

\begin{figure}[!ht] 
\centerline{\psfig{file=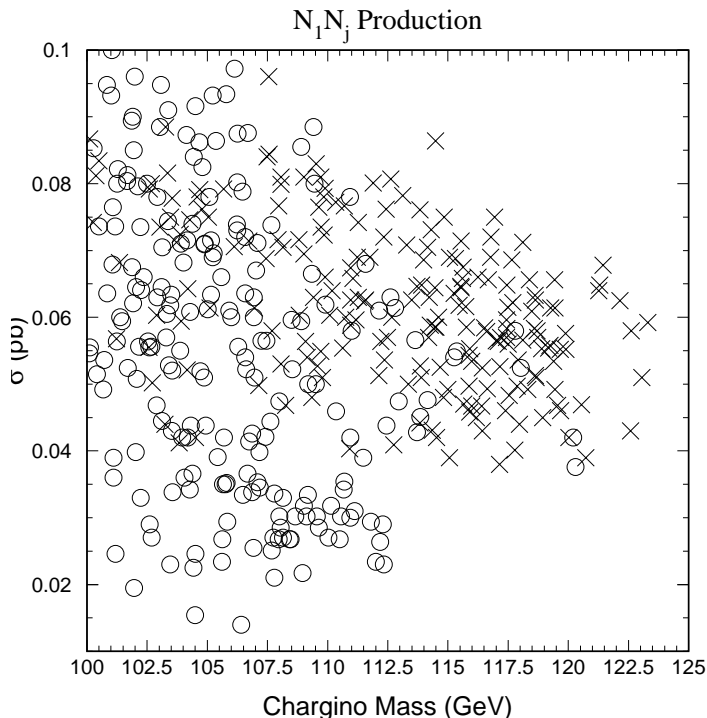,width=10cm}}
\caption{Relation between production cross section
times efficiency and the chargino mass at $\sqrt{s}=200$ GeV
for $\none\n j$ production.}
\label{fig:mc1}
\end{figure}

Some solutions had $\none\nthree$ production as the dominant production
mechanism, while others had only
$\none\ntwo$.  For $\none\nthree$, the solutions clustered
into the regions $110<|\mu|<140$ GeV, $M_2>200$ GeV, $50<M_1<90$ GeV (the exact upper and
lower bounds depend on the sign of $\mu$),
with no real constraint on $\tan\beta$.  For $\none\ntwo$ production, the regions are
$100<|\mu|<135$ GeV, $M_2>250$ GeV, $70<M_1<135$ GeV.
Surprisingly, the $\none$ LSP in all solutions has a large
Bino composition, with $|N_{11}^2|$ in the range $.3-.8$ for $\none\ntwo$ production,
and $.6-.9$ for $\none\nthree$ production.
The requirement that $M_{\cone}>100$ GeV strongly affects the solutions.
Figure~\ref{fig:mc1} shows the relation between the chargino mass and the
observed cross section at $\sqrt{s}=200$ GeV for the dominant $\none\ntwo$
solutions (o) and $\none\nthree$ ones (x).  The chargino is never heavier than
125 GeV, and some solutions predict that the threshold for pair production
is close to the current LEP2 energy.

To summarize, $\none\n j$ production can produce the correct tau excess by providing
an indirect source of tau sleptons.  In essence, the $\none\n j$ process extends
the possible range of tau slepton masses beyond that derived solely from
direct stau production.  The neutralino LSP should also be Bino-like, but mainly
for kinematic reasons.

\leftline{\bf Comment on 3-body Decays:}
The decay $\n j\to\tau^+\tau^-\none$ can have a large branching
fraction in this case only if the tau slepton contribution can
overcome the $Z^*$ one.  This requires a
light tau slepton and large $\tan\beta$.
In the pure Higgsino limit, the equality of
the stau and $Z^*$ couplings requires $g/(2\cos\theta_W)\sim
g m_\tau\tan\beta/(2 M_W \sqrt{2})$ or $\tan\beta=73$, where
we have taken $N_{13}=-N_{14}=N_{23}=N_{24}=1/\sqrt{2}$.
Color factors and additional
couplings increase this value somewhat, which is already beyond
perturbative limits.
Therefore, the mechanism of $\none\n j$ production, followed by
the 3-body decay of $\n j\to \tau^+\tau^-\none$, is not very promising in explaining
the tau excess.

\leftline{\bf Comment on GMSB:}  In GMSB, 
$\none\none$ production is a possible source of tau leptons if
the tau slepton is the NLSP.  However, there are several
difficulties: (1) four tau leptons will be produced in the final
state -- two may be soft, but some of these should still be visible;
(2) one half of the time the two hardest taus will have the same charge;
(3) small $\mu$ is needed, which is difficult to achieve in the
minimal model.  Other production modes, such as $\n i\n j$
or $\cone\cone$, will produce final states with too much structure
from cascade decays.

\subsection{Chargino Pair Production}

Chargino pair production does not appear to be a viable
explanation of a tau excess.
However, it is possible that a slight increase in  collider
energy will cross the kinematic threshold for chargino pair production,
so a brief discussion is in order.

Chargino pair production at LEP proceeds mainly through the
coupling of the chargino pair to an off-shell photon or a
$Z^0$ boson or $t$-channel electron sneutrino exchange.
The $Z\cone\cone$ coupling depends on 
$O^{'L}_{11}=\sin^2\theta_W-|V_{11}|^2-{1\over 2}|V_{12}|^2$
and $O^{'R}_{11}=O^{'L}_{11}(V\to U)$, {\it i.e.} the coupling
is significant whether the chargino is Wino-like, Higgsino-like,
or of a mixed composition.
For this reason,
the chargino production cross is large compared to
other sparticle processes in almost any model.  
The total cross section does depend
on the electron sneutrino mass, and a light electron sneutrino can
significantly reduce the event rate by an order of magnitude from
the heavy sneutrino limit.
The example cross sections shown in Figure~\ref{fig:xsections}, for
the case of $M_{\widetilde{\nu}_e}=100$ GeV, already exhibit a threshold
dependence that is too rapid to explain the data.  In particular,
it is difficult to explain the dip at $\sqrt{s}=200$ GeV, though this
is challenging for all the scenarios considered.
One can consider smaller values for $M_{\widetilde{\nu}_e}$,
but $M_{\widetilde{\nu}_e}<M_{\cone}$ will lead to an electron
excess.

It is important to note that chargino pair production can yield
the correct kinematics.
The decay
$\cone\to\stau_R\nu_\tau$ through a Higgsino component of the
chargino wave function will produce energetic tau leptons from the
$\stau_R$ decay.  The same is true for $\stau_L$, except this does
not require Higgsino components to the chargino, and there will
also be the decay $\cone\to\snut\tau$.  If $M_{\stau_L}>M_{\cone}$, 
then the latter decay may be the dominant one. The
decay $\snut\to\nu\lsp$ is
assumed to complete the chain, 
so that these decays will be essentially invisible.
Large values of $\xvis$ are obtained for $\cone\to\stau\nu_\tau,\stau\to\tau\lsp$
if $M_{\stau}\sim M_{\cone}>.45-.64 M_{\lsp}$.  The decay
$\cone\to\snut\tau$ requires $M_{\snut}<.45-.64 M_{\cone}$ to
generate large $\xvis$.

\section{Theoretical Considerations}

In this section, we analyze the values of the soft SUSY-breaking
parameters that are consistent with the kinematics and event rate
of a tau excess.  One concern is how to arrange for an excess
in only tau leptons.  The phenomenological analysis of the previous
section requires
that a tau slepton is lighter than the selectron.  This is not unreasonable,
since the connection between the selectron and stau masses
depends on theoretical prejudice.  At the high mass scale where
the soft SUSY-breaking parameters of the MSSM are initially induced, the
mass parameters associated with the selectron and stau may only
be loosely related.  For example, if the sleptons have a common mass parameter
at the Planck scale, the evolution of these parameters to the GUT scale, where
extra GUT fields are decoupled, can lead to non-universality.
There are potentially several
different mass scales associated with different physics (the GUT scale physics,
compactification scale, {\it etc.}) which may have different
particle content, thresholds, {\it etc.}  Alternatively, the mass
parameters will be highly correlated in specific models such as
mSUGRA or minimal GMSB.   Here, 
the consequences of some high mass scale assumptions are examined.

Assume that there is a common origin to slepton masses from
universal boundary conditions at some high mass scale.  If there
is essentially no evolution of the mass parameters so that
the selectron and stau mass parameters are the same at the weak scale, then
the presence of off-diagonal terms $m_\tau\widetilde{A}_\tau\equiv
m_\tau(A_\tau-\mu\tan\beta)$ in the stau mass matrix will decrease
the lightest tau slepton mass.  This is the mechanism resulting
in a lighter tau slepton as NLSP in GMSB.
The relations between the lightest tau slepton mass, the
mixing angle $\theta_{\stau}$, and the left- and right-handed
selectron masses are:
\begin{eqnarray} 
M^2_{\tilde{e}_R}-M^2_{\stau_1}={m_\tau|\att|\over \tan\theta_{\stau}}
~~~~~ M^2_{\tilde{e}_L}-M^2_{\stau_1}=m_\tau|\att|{\tan\theta_{\stau}}
~~~~~ \tan 2\theta_{\stau} = {2 m_\tau \att \over 
M^2_{\tilde{e}_R} - M^2_{\tilde{e}_L}},
\end{eqnarray}
using the
convention that $\cos\theta_{\stau}\to 0$ corresponds to
$\stau_1\to\stau_R$.
For $M_{\stau_1}=80$ GeV, $|\mu|=1$ TeV,  
$\tan\beta=50$, $A_\tau=0$, and $\tan\theta_{\stau}=10$,
the selectron mass is 123 GeV.  In all generality, 
it is easy to arrange that tau slepton pairs are produced 
at $\sqrt{s}=200$ GeV
while selectron (and smuon) pairs are not, but
this may not be true within a given theoretical framework.
Note that when selectrons are kinematically accessible, they can be
produced 
with a significantly larger cross section than smuon or tau slepton pairs,
provided that gaugino-like neutralinos are not too heavy.
Since the chargino and neutralino production cross sections at LEP depend
upon the selectron and electron sneutrino masses, the data may not be
compatible with arbitrarily heavy masses.

In models where there is significant renormalization group evolution of
the mass parameters, such as in Supergravity models
with boundary conditions at or near the GUT scale, a large $\tau$
Yukawa coupling can reduce $M_{\stau_L,\stau_R}$ with respect to
$M_{\tilde{e}_L,\tilde{e}_R}$.  This will cause additional splitting
between the lightest tau slepton and selectron mass eigenstates.
Roughly, both effects -- the presence of off-diagonal terms and
the decrease of the diagonal terms -- are of equal importance.

Both of these examples of
generating a large mass difference between the tau slepton and
the selectron
require that $\tan\beta$ is sizeable.  This can 
have other consequences in the sparticle and Higgs sectors, which
will be mentioned later.
An alternative explanation of only a light tau slepton would be the
presence of $D$-terms that are specific to third generation sparticles.
Finally, in models of
``more minimal'' Supersymmetry \cite{Cohen:1996vb}, the sfermions 
associated with
the first two generations are naturally heavy, on the order of $1-10$ TeV.
It is straightforward to have a light tau slepton in this approach.

To make the discussion more concrete, consider the tau slepton production
scenario in a minimal GMSB model.  For large values of $\att\sim-\mu\tan\beta$,
the lightest tau slepton becomes the NLSP, and 
is approximately a right--handed tau slepton interaction state
significantly lighter than the selectron and 
smuon \cite{Dimopoulos:1997yq,Ambrosanio:1997rv}.  
The estimate for the
kinematics is $x_{vis}={1\over 2}$, which is easily
consistent with the data.
The minimal model predicts that
$M_{\tilde{e}_R}\sim 1.1 M_1$ and $M_{\tilde{e}_L}\sim 1.2 M_2$ (ignoring
D-terms), leading to
the relation $M_{\tilde{e}_L}/M_{\tilde{e}_R}\sim 2.2$.  
Two parameters, for example, $M_{\stau_1}$ and $\theta_{\stau}$,
fix the stau and selectron spectrum.
Since a fixed cross section is consistent
with a range of values for $M_{\stau_1}$ and $\theta_{\stau}$, it
is straightforward to check if there are solutions with a heavy
selectron.
After imposing the
restrictions
$M_{\tilde{e}_R}>100$ GeV and $|\att|<50$ TeV, a narrow range of 
values are compatible with a given cross section;
a cross section of
$(.05,.10,.15)$ pb requires $M_{\stau_1}=(88,81,74)$ GeV.  Within
this range, it is always possible for the selectron and smuon to be
beyond the kinematic reach of LEP2, with a reasonable value for $\att$.
The fact that the range of solutions is narrow can be understood from
Eq.~\ref{eq:ratio}, which demonstrates that mixing does not cause
a large variation in the cross section.  
Since $M_{\tilde{e}_R}\sim 1.1 M_1$, the lightest neutralino
may be kinematically accessible at LEP, while the heavier neutralinos and
the chargino are definitely not.  However, $\none$ is expected
to be largely $\widetilde{B}$, so the $\none\none$ production
cross section would not be large.  It is also possible to increase
the neutralino and chargino masses with respect to the slepton ones
by adding multiple representations of messengers.

Consider instead tau slepton production with a neutralino LSP.
In mSUGRA, the approximate relation
$M^2_{\tilde{e}_L}=M^2_{\tilde{e}_R}+.35m_{1\over 2}^2$ holds,
with $M_1\sim .4m_{1\over 2}$.  
Assuming $M_{\none}=M_1=0.4m_{1\over 2}=.45 M_{\stau}$
and repeating the same analysis yields almost identical results as for GMSB. 
Assuming that $M_{\tilde{e}_{R,L}}=M_{\stau_{R,L}}$ is conservative, since RGE
evolution is likely to reduce the stau mass parameters.
One concludes that it is not difficult even within specific models to
arrange for a heavier selectron.
Of course, strict adherence to the mSUGRA mass relations predicts that
the chargino is already kinematically accessible at LEP2, which is not
consistent with the data.  Values of $M_2$ or $|\mu|$ too close to
$M_Z$ will yield too light of a chargino mass.
If tau slepton pair production with a
neutralino LSP is the correct interpretation of the data, then 
some mechanism must split the gaugino mass parameters.  This may be
accomplished with non-universal boundary conditions at the GUT scale,
for example.

The neutralino pair production scenario is also outside of the mSUGRA
framework, since it requires large $M_2$ and $M_1\sim {1\over 2}|\mu|$.
Typical solutions have $M_2/M_1>2.5-3.0$.

There are likely alternative explanations to the tau excess
involving SUSY or not, and we
have not considered them all.
The possibility of R-parity violating decays of $\none$ induced by
an operator $LLE$ can
be rejected: the simplest argument is that each $\none$ decay
must involve two charged leptons, leading to four leptons (some of
which may be tau leptons) in each $\none\none$ event \cite{jack}.  

\section{Predictions for Fermilab}

\subsection{Compatibility with the $\tau\tau\gamma\gamma$ event}
The $e^+e^-\gamma\gamma\met$ event observed by CDF in Run I \cite{Abe:1999ui}
garnered much interest as a candidate for SUSY.  One compelling
explanation was the production and decay of selectron pairs,
with each selectron decaying in the cascade 
$\tilde{e}\to e\ntwo(\to\gamma\none)$ \cite{Ambrosanio:1997wt}.  
However, upon further analysis, the electron candidate seems
to be inconsistent with an electron produced at the production
vertex.  It may be an electron or a pion from the delayed
decay of a tau lepton \cite{henry}.
While this rules out
the selectron explanation, it is consistent with stau pair production.

It would be extremely interesting if the FNAL event and LEP events
had a common explanation.  It is clear that direct tau slepton
pair production {\bf cannot} be the source of the tau excess in
this scenario, since photon pairs should also be observed -- this
also pertains to the GMSB explanation of the FNAL event.
Since neutralino pair production is only an indirect source
of single tau sleptons, 
single, hard photons would be observed in this scenario.
If there is a tau excess
at LEP with a SUSY explanation, then the explanation of
the FNAL event must either
be a highly improbably fluctuation of a Standard Model process,
a gross mismeasurement, or a subtle SUSY effect.

\subsection{General considerations}
There have been several studies on the impact of a light
tau slepton on phenomenology at the Tevatron
\cite{Wells:1998pn,Dutta:1999tt,Muller:1999hw,Baer:1997yi,Lykken:2000kp}.
These have focussed mainly on the modifications to the trilepton
signature.  Based solely on what can be observed at LEP, which may
require only a light tau slepton and the LSP,
not much more
can be predicted for Fermilab without introducing 
theoretical prejudice.

Direct slepton production has never been considered as a promising
avenue for discovering SUSY at hadron colliders.  Considering
only physics backgrounds from real tau leptons, $Z/\gamma^*$,
$WW$ and $ZZ$ production will be the dominant sources.  Since
slepton pair production has a rate of roughly $10-100$ fb$^{-1}$
at the Tevatron, it will be difficult to achieve $S/B<{1\over 10}$.
Establishing a tau slepton signal will clearly be more challenging
than establishing a selectron or smuon one, as is the case at LEP.

If neutralino pair production is the source of the tau excess at LEP,
then SUSY production rates at the Tevatron will be much larger, because
the chargino is light.  $\none\n j$ production cannot be too large,
because the size of the cross section at LEP must be comparable
to slepton pair production.
Chargino pair production will be a much more copious source of
tau lepton pairs, so that $S/B\sim 1/10$ may be achievable.

Since the LSP should be Bino-like, $\ntwo$ and $\cone$ have
similar wave function compositions, so that $\ntwo\cone$ is
a large cross section, leading to the tri-tau signatures 
considered in previous studies.  If $\ntwo$ is Higgsino-like,
then $\ntwo\ntwo$ production will be a source of up to
four tau leptons, though two are likely to be soft.  The
two hard ones can have the same sign.  
It is an interesting
experimental question whether the {\bf single-prong} hadronic
decays of the tau have sufficiently small background to be
part of a like-sign signal.

A significant mass splitting between the selectron and stau masses may
require moderate or large values of $\tan\beta$.
Another consequence of large $\tan\beta$ would be its impact
on Higgs boson phenomenology.
Properties of Higgs bosons
may be significantly altered \cite{CMW}.  In particular, at
large $\tan\beta$, a sbottom-bottom-gluino loop can shift the
bottom quark Yukawa from its tree level value, possibly leading
to enhanced light Higgs boson decays to tau, $W$ and photon pairs.  
As is usual in
large $\tan\beta$ models, the associated production of Higgs bosons
with bottom quark pairs would be a promising avenue for 
discovering one or more Higgs bosons.
If large $\tan\beta$ is indeed a necessary component
of the explanation, then the bottom squark
may also be light, leading to a $b\bar b\slashchar{E}_T$ signature
\cite{Demina:1999ty}.  The signature is unlikely to be modified
by the presence of a light chargino and tau slepton.  Of course,
in the neutralino pair scenario, the bottom quark cannot be so light
that it influences the neutralino decays.

Furthermore, there may be a light top squark, though this is not
required by the results of our phenomenological analysis of
the tau excess.  If we
exclude the possibility of a top squark so light that
$\widetilde t\to c\none$ is relevant, then the decay
$\widetilde t\to b\cone$ will occur, with $\cone\to\stau\nu_\tau$.
Stop pair production can then lead to a sample of top-like
events with an enrichment of tau leptons.  Assuming a top squark
mass equal to the top mass, and the approximate relation that 
top squark production is ${1\over 10}$ of top quark production,
then there would be roughly 25 events containing $e$ or $\mu$ + one hadronic
tau decay in $100$ pb$^{-1}$ of data.  If the detection efficiency is
several percent, then this would be compatible with the CDF
data \cite{Abe:1997uk}.

For some time, there has been indirect evidence that the gluino
is light \cite{Kane:1999im}.  Once we give up on the idea of a 
universal boundary condition for the gaugino masses, as is
required by the results of the phenomenological analysis, there is
no compelling reason to believe that the gluino is significantly 
heavier than the LSP.
Pair production of gluino pairs
or the associated production of gluinos and charginos are
a source of like-sign charginos, which decay into like-sign
tau leptons.  The branching ratio for a pair 
$e$ or $\mu$ leptons from $\tau$ pair decay
is about 12\%, but some of the leptons
make be fairly soft.  Clearly, the use of hadronic tau decays
will improve the experimental sensitivity.

To summarize, a tau excess at LEP may translate directly into
a tau excess at Tevatron if a number of sparticles are light.
Some likely light candidates are the gluino and the bottom and
top squarks.  The reasons that each of these are light, however,
may have different physics origins that are not directly related
to the existence of a light tau slepton.

\section{Cosmological consequence of a light tau slepton}

One interesting consequence of a stable LSP is a potential
solution to all or part of the dark matter problem.
A consistent explanation of the tau
excess in models with a neutralino LSP led us to 
consider a right-handed tau slepton and a Bino-like LSP.
For such a case, the relic abundance
is given by \cite{Wells:1998ci}:
\begin{eqnarray}
\Omega h^2 = {M_{\widetilde\ell_R}^2 \over M_0^2 \sqrt{N_F} 
{1\over 3}N_{\widetilde\ell_R}}
r^{-2} {(1+r^2)^4 \over 1+r^4}, r= {M_{\none}\over M_{\widetilde\ell_R}},
\end{eqnarray}
where $M_0\simeq 460$ GeV, $N_F=606/8$ for the mass range under
consideration, and $N_{\widetilde\ell_R}$ counts the effective
number of light right-handed sleptons.  $N_{\widetilde\ell_R}=3$ 
for three mass degenerate right-handed sleptons, and
$=\sin^4\theta_\tau$ when a light stau dominates with 
mixing angle $\theta_{\stau}$.  Based on the kinematic analysis of
the data, the range $.45<r<.64$ can be compatible with
the observed $\xvis$.  Fixing $M_{\stau_R}=90$ GeV and $N_{\widetilde\ell_R}=1$
leads to the range $.10<\Omega h^2 < .12$.  Larger values can be  obtained
if the purity of the stau eigenstate is reduced, smaller ones if the
selectron and smuon are also light (but heavier than the stau).
Effects such as
co-annihilation or $Z/h$ pole can only diminish the relic abundance.
In the neutralino pair production scenario, the Bino composition
of the LSP$\equiv |N_{11}|^2$ is approximately $1\over 2$, leading to a relic abundance
almost twice as large.

In the approximations made in the kinematic analyses above, 
the visible energy fraction is consistently $x_{vis}=1/2(1-r^2)$.
For the case of neutralino pair production and decay, this
requires that the contribution from the $\n j$ decay and
from the tau slepton decay are similar in magnitude, which
is reasonable.
Using this relation, it is possible to re-express the equation above in
terms of a collider observable.  Using the previous example
but without fixing $r$, one  finds:
\begin{eqnarray}
\Omega h^2\simeq {.1 \over 1-2 x_{vis}} {(1-x_{vis})^4 \over 1-2x_{vis}+2x_{vis}^2}.
\end{eqnarray}
This relation indicates that
$\Omega h^2\simeq .1$ is expected unless the mass splitting
between the NLSP and LSP is quite large, leading to $E_{vis}/\sqrt{s}\to
{1\over 2}$.

\section{Discussion and Conclusions}

This study has considered the interpretation of a tau excess at
LEP within the MSSM.
After an analysis of the size of the cross section, the kinematics,
and the absence of other signatures, we conclude that the most likely
explanation of a tau  excess at LEP is neutralino or
tau slepton pair production. 
Here, we summarize
the various scenarios considered to generate an excess of
only tau leptons with a fairly large fraction of visible energy.
Direct tau slepton pair production may be the dominant
source of a tau excess.
The cross section for $M_{\stau}=80$ GeV is in agreement with the data.
Unless $\tan\beta$ is small, a right-handed tau slepton is preferred.
A gravitino LSP can easily accommodate the kinematics, as well as 
a neutralino LSP with mass less than about $.45-.64M_{\stau}$.  A neutralino
LSP should be Bino-like to be consistent with the invisible width
of the $Z$ boson.  For the case of a neutralino LSP, many motivated
models would predict other light sparticles within reach of LEP
searches, but there is no such prediction from our phenomenological
analysis.

Alternatively, the associated 
production of neutralino pairs $\none\ntwo$ and/or
$\none\nthree$, followed by two-body decays of the
heavy neutralinos to a tau and stau, 
may be  the dominant source of a tau excess.
The tau slepton should
have a mass that is separated enough from both neutralino
masses so that the tau leptons are fairly energetic.  The
allowed values of $M_{\stau}$ extend beyond those compatible
with solely direct stau pair production.
Despite the fact that the production process proceeds through
Higgsino components of the neutralinos, 
the neutralino LSP has a large Bino component to its wave function.
This is fixed by the large mass splitting needed between the
two neutralinos to generate the correct kinematics.
The chargino is predicted to be light, though not yet kinematically
accessible at LEP, or it would influence the neutralino decays or
produce its own excess.

Finally, chargino pair production may be the dominant source of
a tau excess, but the cross section turns on rapidly above the
kinematic threshold, and is not likely to explain all data.
It is straightforward to satisfy the kinematic constraints of
the observed excess if a stau or tau sneutrino is lighter than
the chargino, so chargino pair production will be readily observable
once the kinematic threshold is crossed.

Tau slepton pair production with a gravitino LSP falls within
the minimal GMSB framework.  The case of a neutralino LSP,
in conjunction with tau slepton and/or neutralino pair production,
requires a deviation from the usual mSUGRA relations between
gaugino masses.  Neutralino pair production requires $M_1\sim
{1\over 2}|\mu|$ and $M_2/M_1>2.5$.  In all cases, the theoretical
framework should predict that the selectron and smuon are heavier
than the stau.  Large values of $\tan\beta$ may be necessary
to accomplish this.

At LEP, a marginal increase in
energy will not substantially increase the tau
slepton or neutralino cross sections,
and the expected amount of data cannot establish more than a $4\sigma$ 
effect \cite{favara}.
However, in either case, but particularly in the neutralino pair scenario,
a kinematic threshold for a new process may well be nearby.  If the
threshold for chargino
pair production is crossed, the tau signal will be enhanced considerably.
In such an interpretation, the neutralino can still be a viable CDM candidate.

The situation at hadron colliders, particularly the Tevatron,
depends on how many sparticles besides the tau slepton
and LSP are light.  The production of neutralinos
depends on their wave functions, but charginos
will likely be produced copiously.  The mSUGRA trilepton signature
becomes more specifically a tri-tau excess.
Top squark decays to bottom and
a chargino can lead to top-like events with an excess of tau leptons.
Gluino pair or gluino-chargino production can lead to like-sign
tau events, which, in turn, can yield like-sign $e$'s or $\mu$'s.
Knowing some sparticle masses from LEP measurements would greatly
enhance detection prospects.

\centerline{\bf Acknowledgements}
We thank M.~Gruenewald, J.F.~Gunion, G.~Kane and J.D.~Wells
for useful comments and criticisms.


\end{document}